\documentclass[runningheads]{llncs}
\usepackage{amsmath}
\usepackage{booktabs}
\usepackage{multirow}
\usepackage[T1]{fontenc}
\usepackage{graphicx}
\usepackage{marvosym}
\usepackage{xspace}
\usepackage[table]{xcolor}
\newcommand{\Name}{\texttt{WeSCE}\xspace}
\definecolor{tabheader}{RGB}{200,220,245}
\definecolor{tabalt}{RGB}{240,245,252}
\begin{document}
\title{WeSCE: A Benchmark for Measuring Security Drift in LLM-Driven Code Editing}
%
%
\author{Zhiyu Zhang \and Tingyue Wen \and Senke Sun \and Dengxiang Liang \and\\Enhao Huang\textsuperscript{(\Letter)}\thanks{Corresponding author: \email{huangenhao@zju.edu.cn}}}
\authorrunning{Z. Zhang et al.}
\institute{Zhejiang University, Hangzhou, China\\
\email{\{zy-zhang,tingyuewen,saski,dengxiang,huangenhao\}@zju.edu.cn}}

\maketitle              
\begin{abstract}
Large language models (LLMs) are increasingly used in software development for code generation and editing, yet how security properties evolve under such model-driven code transformations remains poorly understood. Existing benchmarks primarily evaluate functional correctness or discrete vulnerability detection, but do not explicitly characterize the \textit{security drift} induced by single-step code edits.

In this work, we introduce \Name, a benchmark for quantifying \textit{security drift in code editing under weak-security constraints}, where tasks specify only functional objectives without explicit security requirements. \Name consists of 400 executable programs derived from real-world code, covering feature addition, feature removal, bug fixing, and refactoring. To quantify security drift, we propose a continuous risk representation that aggregates heterogeneous vulnerability signals through a unified formulation, and define drift measures capturing changes in overall risk, worst-case severity, and vulnerability distribution under code transformations, providing a multi-scale view of security spanning average-case behavior to worst-case emphasis.

\end{abstract}
\section{Introduction}

Large language models (LLMs) and LLM-based agents are increasingly integrated into software development workflows, enabling code generation and editing at scale across tasks such as feature development, bug fixing, and refactoring~\cite{huang2024agentcoder,Wu2023llmvul}. As a growing share of production code is influenced by LLM-driven transformations, understanding how these transformations affect code security has become a pressing concern. Recent studies~\cite{vechev2023security,copilotsecurity2022} show that LLM-generated code can introduce unsafe patterns, motivating evaluation that goes beyond functional correctness.

In response, substantial effort has been devoted to benchmarking LLM code capabilities. Existing benchmarks primarily assess functional correctness~\cite{Chen2021HumanEval,Austin2021MBPP,Liu2023evalplus} or execution quality~\cite{jain2024livecodebench}. Security-oriented benchmarks further consider vulnerability detection or repair~\cite{Peng2025CWEval,chen2026seccodebench,pellew2026realvuln}, but typically evaluate static code snapshots or tasks with explicit security objectives.

However, we argue that this evaluation paradigm overlooks a practically important scenario. In real-world development, most code edits are driven by functional requirements---adding features, fixing bugs, or refactoring---without explicit security instructions. Under such \textit{weak-security constraints}, security properties may change implicitly as a side effect of editing, yet no existing benchmark explicitly characterizes this \textit{security drift}: how vulnerability signals evolve under non-security-driven code transformations.

Moreover, current security evaluations treat vulnerability as a discrete, binary property. A program is labeled either ``vulnerable'' or ``safe,'' but this labeling fails to capture important nuances. For instance, a refactoring may remove one vulnerability while introducing a different one elsewhere; both the original and edited versions would receive the same ``vulnerable'' label, obscuring the redistribution and relative severity of risk. To systematically track how security changes under code editing, a continuous and multi-dimensional representation is needed.

\begin{figure}[htbp]
\centering
\makebox[\textwidth][c]{\includegraphics[width=1\textwidth]{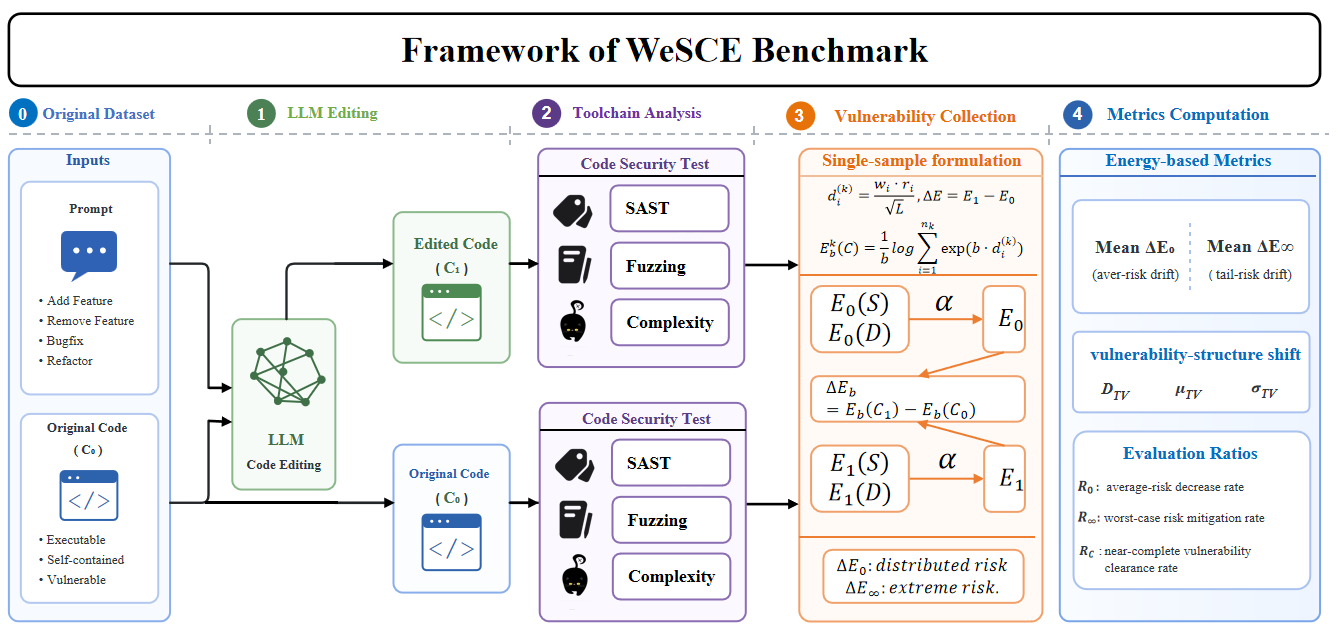}}
\caption{Overview of the \Name evaluation framework. An LLM edits a program under weak-security constraints (i.e., functional-only instructions). Both the original and edited code undergo static analysis (SAST), dynamic fuzzing, and complexity measurement. Vulnerability signals are then aggregated into continuous risk scores $E_b$ via LogSumExp, and security drift metrics ($\overline{\Delta E_0}$, $\overline{\Delta E_\infty}$, $D_{\mathrm{TV}}$, $R_0$, $R_\infty$, $R_c$) are computed by comparing pre- and post-edit risk representations.}
\label{fig_framework}
\end{figure}
\vspace{-2mm}
To address these gaps, we introduce \Name (\textbf{We}ak-\textbf{S}ecurity \textbf{C}ode \textbf{E}diting), a benchmark for quantifying security drift under weak-security constraints. The key insight behind \Name is that security should be modeled as a continuous attribute of programs, enabling fine-grained tracking of how risk magnitude, worst-case severity, and vulnerability distribution change after each edit. Concretely, we propose a risk representation based on LogSumExp aggregation that smoothly interpolates between average-case and worst-case sensitivity, and define drift measures that capture directional changes in this representation. The benchmark comprises 400 executable programs derived from real-world code on GitHub and Real-Vuln-Benchmark~\cite{pellew2026realvuln}, covering four transformation types: feature addition, feature removal, bug fixing, and refactoring. All programs are validated for executability and deduplicated via CodeBERT~\cite{feng2020codebert_emnlp}.

We evaluate eight state-of-the-art LLMs on \Name. Results reveal a consistent model stratification: stronger models (e.g., Opus 4.6, GPT-5.4) achieve risk reduction in over 80\% of samples, while weaker models fall below 60\%. Across all models, worst-case vulnerability mitigation is more pronounced than average-case reduction, suggesting implicit sensitivity to high-severity patterns. However, even the best model achieves complete clearance in only about half of samples, indicating that current LLMs primarily provide partial mitigation rather than full security resolution under weak-security constraints.

In summary, our main contributions are as follows:
\begin{enumerate}
\item We formalize the problem of \textit{security drift under weak-security code editing}, a practically important but previously uncharacterized scenario in which code transformations are driven by functional objectives and security changes occur as an implicit side effect.
\item We construct \Name, a benchmark of 400 executable programs with four editing task types, designed for controlled analysis of security drift under realistic development conditions.
\item We propose a drift-oriented evaluation framework with continuous risk scores and multi-scale drift measures that capture average risk change, worst-case severity change, and vulnerability redistribution.
\item We evaluate eight LLMs and provide systematic findings on model stratification, worst-case mitigation bias, and the relationship between transformation depth and security drift.
\end{enumerate}

\section{Related Work}

\subsection{Benchmarks for LLM-Based Code Generation}

Benchmarking LLMs for code generation spans multiple dimensions. Early benchmarks such as HumanEval~\cite{Chen2021HumanEval} and MBPP~\cite{Austin2021MBPP} focus on unit-test-based correctness, while EvalPlus~\cite{Liu2023evalplus} improves test coverage. Other efforts address contamination via dynamic evaluation~\cite{jain2024livecodebench}, and efficiency-oriented benchmarks~\cite{Huang2024effibench,Liu2024evalperf} incorporate performance profiling.

Security-oriented benchmarks extend these evaluations to vulnerability detection. SecCodeBench~\cite{chen2026seccodebench} provides generation and repair tasks with CWE-level security verification, and MT-Sec~\cite{rawal2025benchmarking} reveals that security and correctness degrade over multi-turn interactions. However, these benchmarks still treat security as a discrete outcome (vulnerable or safe) and do not characterize how vulnerability signals change under code transformations.

\subsection{Code Editing and Security under Transformations}

Recent work has explored more realistic coding scenarios, including automated program repair~\cite{campos2026exploring} and multi-agent code generation~\cite{huang2024agentcoder,Wu2023llmvul,ase_benchmark}. These efforts often incorporate execution-based validation or focus on fixing known vulnerabilities. For example, PatchEval~\cite{wei2025patcheval} provides sandboxed environments for testing fixes, and PatchLM~\cite{Guru2025Generating} is trained to generate security patches from CVE-associated commit hunks.

However, these approaches typically assume an explicit security or bug-fix objective, which does not reflect routine development practice. In practice, most code changes are driven by functional requirements~\cite{Guru2025Generating}, and any security impact is incidental. No existing approach explicitly models how security properties evolve under such non-security-driven code transformations.

\subsection{Security Assessment and Representation}

Existing security evaluation benchmarks~\cite{ase_benchmark,chen2026seccodebench,pathak2025dualguage} typically reduce security to a static binary label, classifying code as vulnerable or safe under fixed CWE categories. While this labeling can flag known flaws, it cannot capture that a single edit may simultaneously remove one vulnerability and introduce another, leaving the overall security posture changed but unlabeled.

Recent efforts attempt to combine functionality and security into unified metrics. CWEval~\cite{Peng2025CWEval} tests functional correctness alongside vulnerability checks, and RealSec-bench~\cite{wang2026realsecbench} introduces a composite SecurePass metric. However, both still produce coarse pass/fail outcomes and do not model how security evolves during code transformations.

In contrast, our work models security as a continuous attribute and introduces drift measures that explicitly track fine-grained changes in risk magnitude, worst-case severity, and vulnerability distribution after each edit. This continuous representation also supports extensibility: as new vulnerability types emerge, additional signals can be incorporated without modifying the aggregation framework.

\section{Framework and Method of \Name}

\Name is designed around three principles: (1) tasks specify only functional objectives without security cues, operating on fully executable programs; (2) security is characterized through both static vulnerability signals and dynamic execution risks as complementary indicators; and (3) heterogeneous signals are aggregated into a continuous risk representation, enabling fine-grained tracking of how risk magnitude, worst-case severity, and vulnerability distribution change after each edit.

\subsection{\Name Data Source}

\begin{figure}[htbp]
\centering
\makebox[1\textwidth][c]{\includegraphics[width=1\textwidth]{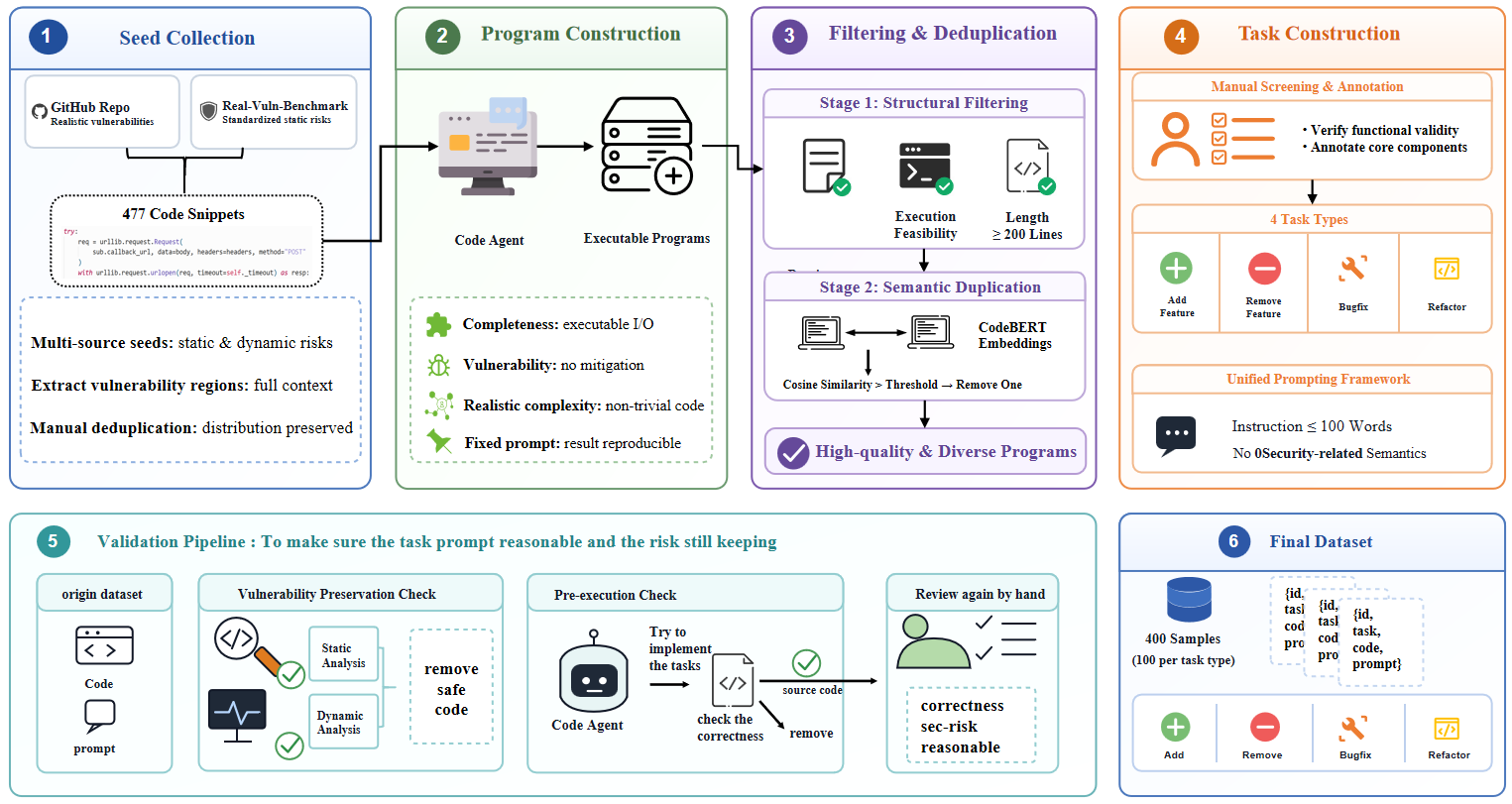}}
\caption{The \Name dataset construction pipeline. Vulnerability seeds are collected from GitHub and Real-Vuln-Benchmark, expanded into fully executable programs via a code agent with fixed prompts, filtered by syntax and executability constraints, and deduplicated using CodeBERT embeddings. Four editing tasks (ADD, REMOVE, FIX, REFACTOR) are constructed with functional-only instructions. A final validation pipeline verifies program executability, vulnerability preservation, and task prompt correctness.}
\label{fig_data}
\end{figure}
\vspace{-2mm}
\paragraph{Seed Collection.}
We collect vulnerability seeds from GitHub and Real-Vuln-Benchmark~\cite{pellew2026realvuln}, yielding 477 samples, including 143 from Real-Vuln-Benchmark and 334 from GitHub. These seeds cover static issues (e.g., unsafe APIs, injection patterns) and dynamic vulnerabilities (e.g., buffer overflows, runtime errors). Relevant code regions are extracted from issue descriptions or annotations, and near-duplicates are removed using CodeBERT embedding similarity and manual inspection.

\paragraph{Program Construction.}
We expand seeds into executable programs using the Claude Code agent~\cite{Claude-Code} with a fixed prompt (released for reproducibility). For fairness analysis, we additionally construct an alternative expansion pipeline using Cursor with DeepSeek-V4, applied to a subset of 40 seeds. For each seed, multiple candidates are sampled under identical decoding settings, and valid programs are selected based on four constraints: executable interfaces, preservation of vulnerability-related behavior (verified via static and dynamic analysis), realistic control/data-flow expansion, and a minimum size of 200 LOC with multi-function structure.

The final dataset used in main experiments is generated using the Claude-based pipeline to ensure consistency across samples.

\paragraph{Filtering and Task Construction.}
Invalid samples are filtered by syntax, executability, and length constraints, and remaining programs are deduplicated using CodeBERT embeddings~\cite{feng2020codebert_emnlp}. We construct four editing tasks: \textit{add feature}, \textit{remove feature}, \textit{bugfix}, and \textit{refactor}, using a unified instruction template ($\leq$100 words) without explicit security cues.

\paragraph{Quality Control and Statistics.}
Program validity is verified via static analysis and dynamic execution, with additional manual inspection on a subset of 40 randomly sampled programs (out of 400) to ensure correctness and vulnerability consistency. The final dataset contains 400 samples evenly distributed across four task types, each consisting of an executable program and a validated instruction.

\subsection{Assessment Design}
\label{sec:assessment}

\paragraph{Single-sample formulation.}

Given a program $C$, let $r_i(C)$ denote the number of detected vulnerabilities of type $i$ in category $k \in \{s,d\}$ (static or dynamic), $w_i$ the corresponding severity weight, and $L(C)$ the number of logical lines of code. We define the normalized vulnerability signal as
\[
d_i^{(k)}(C) = \frac{w_i \cdot r_i(C)}{\sqrt{L(C)}}.
\]

The $\sqrt{L}$ normalization reduces program-size bias: larger programs tend to contain more vulnerability signals due to increased code surface, and vulnerability density often grows sublinearly with program size. As shown in Table~\ref{tab:scaling_cv}, $\sqrt{L}$ normalization yields the lowest coefficient of variation (CV = 0.41) compared to no normalization (0.82), linear scaling (0.74), and logarithmic scaling (0.63), confirming its effectiveness in stabilizing vulnerability density across programs of varying sizes.

\begin{table}[htbp]
\centering
\caption{Stability comparison of different normalization strategies for vulnerability density estimation. CV denotes the coefficient of variation across programs of varying sizes; lower values indicate more stable density estimates.}
\label{tab:scaling_cv}
\begin{tabular}{l c}
\toprule
\rowcolor{tabheader}
\textbf{Normalization Method} & \textbf{CV of Density} \\
\midrule
\rowcolor{tabalt}
No normalization ($r_i$) & 0.82 \\
$1/L$ scaling & 0.74 \\
\rowcolor{tabalt}
$\log L$ scaling & 0.63 \\
$\sqrt{L}$ scaling (ours) & \textbf{0.41} \\
\bottomrule
\end{tabular}
\end{table}
\vspace{-2mm}
\paragraph{Risk aggregation via LogSumExp.}

To aggregate heterogeneous vulnerability signals into a single risk score, we employ a LogSumExp operator, which provides a smooth interpolation between average-case and worst-case aggregation. Specifically, for each category $k$, let $n_k$ denote the number of vulnerability types. We define
\[
E_b^{(k)}(C) = \frac{1}{b} \log \frac{1}{n_k}\sum_{i=1}^{n_k} \exp\!\left(b \, d_i^{(k)}(C)\right),
\]
where $b > 0$ is a sensitivity parameter controlling the emphasis on high-severity signals. The overall risk score combines static and dynamic components:
\[
E_b(C) = \alpha \, E_b^{(s)}(C) + (1 - \alpha) \, E_b^{(d)}(C),
\]
where $\alpha \in [0,1]$ balances the two categories.

Intuitively, when $b$ is small, $E_b^{(k)}$ approximates the arithmetic mean of signals, treating all vulnerability types equally; when $b$ is large, it approaches the maximum signal, emphasizing the most severe vulnerability. We denote these two limiting regimes by
\[
E_0(C) = \lim_{b \to 0} E_b(C),
\quad
E_\infty(C) = \lim_{b \to \infty} E_b(C).
\]

\paragraph{Security drift measures.}

To quantify how a code transformation from $C_0$ to $C_1$ affects security, we define the drift in average-sensitive and worst-case-sensitive risk:
\[
\Delta E_0 = E_0(C_1) - E_0(C_0),
\quad
\Delta E_\infty = E_\infty(C_1) - E_\infty(C_0).
\]
A negative $\Delta E$ indicates risk reduction, while a positive value indicates risk amplification.

For distributional analysis, we aggregate signals across both categories and normalize them into a vulnerability profile:
\[
p_i(C) = \frac{\sum_{k} d_i^{(k)}(C)}{\sum_{j}\sum_{k} d_j^{(k)}(C)},
\]
and measure structural changes using total variation (TV) distance:
\[
D_{\mathrm{TV}}(C_0, C_1) = \frac{1}{2} \sum_i \left| p_i(C_0) - p_i(C_1) \right|.
\]

\paragraph{LLM-level (batch) evaluation.}

Given a task set $\mathcal{T}$, we compute aggregated statistics over all samples:
\[
\overline{\Delta E_0} = \frac{1}{|\mathcal{T}|} \sum_{t} \Delta E_0^{(t)},
\quad
\overline{\Delta E_\infty} = \frac{1}{|\mathcal{T}|} \sum_{t} \Delta E_\infty^{(t)},
\]
\[
\mu_{\mathrm{TV}} = \frac{1}{|\mathcal{T}|} \sum_{t} D_{\mathrm{TV}}^{(t)},
\quad
\sigma_{\mathrm{TV}} = \sqrt{\frac{1}{|\mathcal{T}|} \sum_{t} \bigl(D_{\mathrm{TV}}^{(t)} - \mu_{\mathrm{TV}}\bigr)^2}.
\]

We further define three rate-based metrics. The risk reduction rates $R_0$ and $R_\infty$ measure the fraction of samples where editing reduces average-sensitive and worst-case-sensitive risk, respectively:
\[
R_0 = \frac{1}{|\mathcal{T}|} \sum_{t} \mathbf{1}\!\left(\Delta E_0^{(t)} < 0\right),
\quad
R_\infty = \frac{1}{|\mathcal{T}|} \sum_{t} \mathbf{1}\!\left(\Delta E_\infty^{(t)} < 0\right).
\]
The complete clearance rate $R_c$ measures the fraction of samples with near-zero residual vulnerability after editing:
\[
R_c = \frac{1}{|\mathcal{T}|} \sum_{t} \mathbf{1}\!\left(E_0\!\left(C_1^{(t)}\right) + E_\infty\!\left(C_1^{(t)}\right) \le \epsilon\right),
\]
where $\mathbf{1}(\cdot)$ is the indicator function and $\epsilon$ is a small threshold.

Together, these metrics capture average risk reduction ($\overline{\Delta E_0}$), worst-case risk reduction ($\overline{\Delta E_\infty}$), structural redistribution ($D_{\mathrm{TV}}$), and complete risk elimination ($R_c$) under LLM-driven code editing.

\section{Experiments}

We evaluate the ability of large language models (LLMs) to perform secure code editing under weak-security constraints using the proposed \Name benchmark. We first describe the experimental setup, followed by overall and task-level analyses of security drift.

\begin{table}[htbp]
\centering
\caption{Mean and standard deviation of total variation (TV) distance $D_{\mathrm{TV}}$ for each model across four editing tasks. Higher $D_{\mathrm{TV}}$ indicates larger distributional shifts in vulnerability profiles after editing.}
\label{tab1}
\small
\begin{tabular}{lcccccccccc}
\toprule
\rowcolor{tabheader}
\textbf{Model} &
\multicolumn{2}{c}{\textbf{ADD}} & \multicolumn{2}{c}{\textbf{REMOVE}} & \multicolumn{2}{c}{\textbf{FIX}} & \multicolumn{2}{c}{\textbf{REFACTOR}} & \multicolumn{2}{c}{\textbf{Total}} \\
\rowcolor{tabheader}
& Mean & SD & Mean & SD & Mean & SD & Mean & SD & Mean & SD \\
\midrule
\rowcolor{tabalt}
Opus 4.6               & 0.517 & 0.317 & 0.409 & 0.263 & 0.725 & 0.241 & 0.866 & 0.341 & 0.629 & 0.343 \\
GPT-5.4                & 0.449 & 0.391 & 0.334 & 0.359 & 0.743 & 0.209 & 0.834 & 0.219 & 0.590 & 0.368 \\
\rowcolor{tabalt}
Sonnet 4               & 0.613 & 0.305 & 0.384 & 0.228 & 0.688 & 0.190 & 0.815 & 0.267 & 0.625 & 0.297 \\
Deepseek-v4-flash      & 0.501 & 0.289 & 0.420 & 0.432 & 0.713 & 0.128 & 0.772 & 0.201 & 0.602 & 0.321 \\
\rowcolor{tabalt}
Kimi k2.5              & 0.533 & 0.362 & 0.524 & 0.320 & 0.607 & 0.213 & 0.766 & 0.190 & 0.607 & 0.298 \\
Haiku 4.5              & 0.431 & 0.362 & 0.599 & 0.394 & 0.598 & 0.359 & 0.677 & 0.341 & 0.576 & 0.376 \\
\rowcolor{tabalt}
Doubao-1.5-pro 32k     & 0.320 & 0.332 & 0.391 & 0.373 & 0.431 & 0.169 & 0.514 & 0.170 & 0.414 & 0.286 \\
GLM-4-32B              & 0.340 & 0.509 & 0.442 & 0.402 & 0.434 & 0.400 & 0.603 & 0.549 & 0.455 & 0.480 \\
\bottomrule
\end{tabular}
\end{table}

\subsection{Experimental Setup}

\paragraph{Models.}
We evaluate eight state-of-the-art LLMs spanning different capability tiers: Opus 4.6~\cite{Opus-4.6}, GPT-5.4~\cite{GPT-5.4}, Sonnet 4~\cite{Sonnet-4}, Deepseek-v4-flash~\cite{DeepSeek-V4}, Kimi k2.5~\cite{Kimi-k2.5}, Haiku 4.5~\cite{Haiku-4.5}, Doubao-1.5-pro (32k)~\cite{Doubao-1.5-pro-32k}, and GLM-4-32B~\cite{GLM-4-32B}. All models are evaluated under a unified decoding configuration (top-$p$ = 0.9, top-$k$ = 20) to ensure consistent output diversity and stability.

\paragraph{Security analysis tools.}
We use CodeQL and Bandit for static analysis and Atheris for dynamic fuzzing-based analysis. Atheris is executed with a fixed 90-second fuzzing budget per program to ensure consistent coverage. To mitigate false positives from static tools, we apply cross-tool consistency checks. We additionally conduct manual validation on a random subset of 40 programs, confirming 94.6\% precision and an estimated 97.2\% recall (with bootstrap confidence intervals) among detected vulnerability instances.

\paragraph{Evaluation parameters.}
We set the static-dynamic weighting to $\alpha = 0.3$, assigning greater weight to the dynamic component to emphasize execution-level vulnerability signals that complement static analysis. For the LogSumExp sensitivity parameter, we test $b = 10^{-3}$ (average-dominant) and $b = 10^{3}$ (worst-case-dominant) to evaluate drift under both aggregation regimes. The clearance threshold is set to $\epsilon = 10^{-2}$, filtering negligible residual signals while preserving non-trivial vulnerabilities.

\paragraph{Evaluation metrics.}
We report six metrics defined in Section~\ref{sec:assessment}: average risk drift ($\overline{\Delta E_0}$), worst-case risk drift ($\overline{\Delta E_\infty}$), TV distance ($\mu_{\mathrm{TV}}$, $\sigma_{\mathrm{TV}}$), risk reduction rates ($R_0$, $R_\infty$), and complete clearance rate ($R_c$). We additionally report structural statistics (LOC and cyclomatic complexity) using the radon toolchain for descriptive analysis.

\begin{figure}[htbp]
\centering
\includegraphics[width=\textwidth]{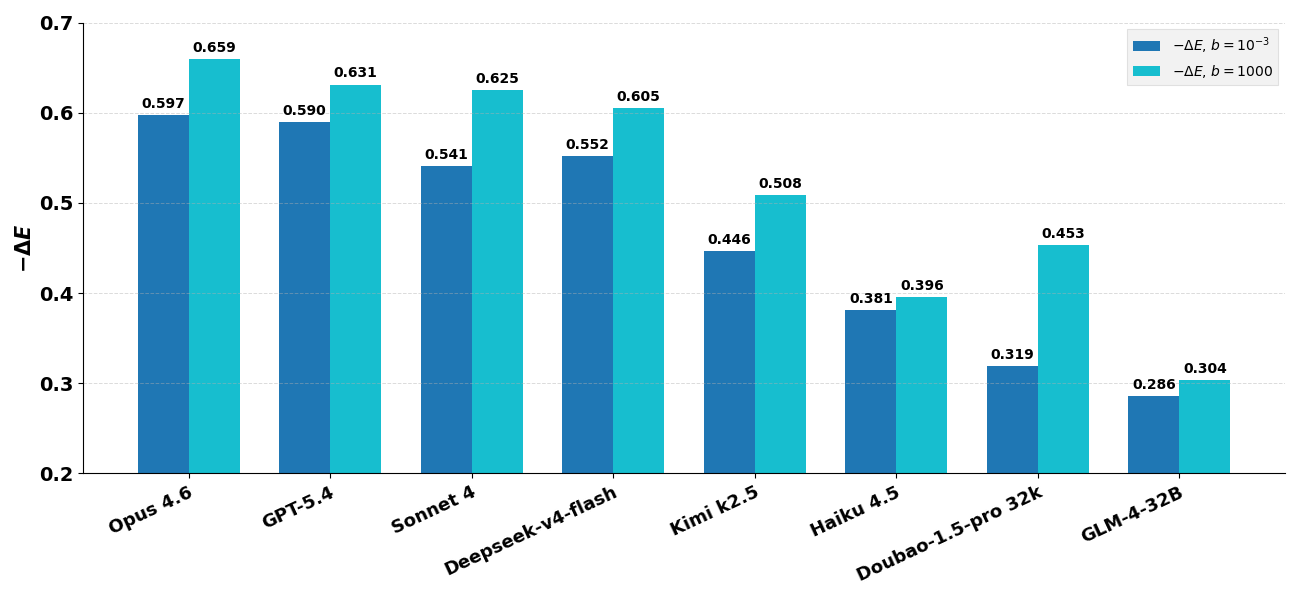}
\caption{Overall risk reduction magnitude $-\Delta E$ (plotted as positive values) for eight LLMs under average-sensitive ($b{=}10^{-3}$, dark blue) and worst-case-sensitive ($b{=}10^{3}$, light blue) aggregation regimes. Taller bars indicate larger security improvements after editing. Stronger models (left) consistently achieve greater reductions, with worst-case reductions exceeding average-case reductions for all models.}
\label{fig_overall_perf}
\end{figure}
\vspace{-2mm}

\subsection{Overall Performance Analysis}

We analyze model behavior on \Name from three complementary perspectives: risk reduction magnitude, worst-case mitigation tendency, and structural redistribution of vulnerabilities.

\paragraph{Model stratification.}
As shown in Table~\ref{tab2}, models exhibit a clear three-tier hierarchy. The top tier---Opus 4.6 ($R_\infty{=}82.25\%$, $R_c{=}51.25\%$), GPT-5.4 ($R_\infty{=}80.00\%$, $R_c{=}50.50\%$), and Sonnet 4 ($R_\infty{=}79.25\%$, $R_c{=}50.50\%$)---achieves risk reduction in over 79\% of samples and complete clearance in roughly half. The mid-tier models (Deepseek-v4-flash, Kimi k2.5) achieve $R_\infty$ between 71--75\% but lower $R_c$ ($\leq$49\%). The lower tier (Haiku 4.5, Doubao-1.5-pro, GLM-4-32B) falls below 66\% in $R_\infty$, with $R_c$ as low as 31.50\% for GLM-4-32B. This stratification is consistent with Fig.~\ref{fig_overall_perf}, where the top-tier models show the largest negative $\overline{\Delta E_0}$ and $\overline{\Delta E_\infty}$.

Notably, the gap between top- and bottom-tier models is substantial: Opus 4.6 outperforms GLM-4-32B by 23.50 percentage points in $R_\infty$ and 19.75 points in $R_c$, indicating that model capability plays a significant role in incidental security improvement.

\paragraph{Worst-case mitigation bias.}
Across all models, we observe a consistent pattern:
\[
|\overline{\Delta E_{\infty}}| > |\overline{\Delta E_0}|,
\]
indicating that worst-case vulnerability reductions are more pronounced than average-sensitive ones. We attribute this to the nature of high-severity vulnerabilities: they tend to involve explicit insecure patterns (e.g., injection, unsafe deserialization) that are more recognizable and thus more likely to be altered during editing. This suggests that LLMs possess implicit security awareness targeting salient vulnerability patterns.

\paragraph{Incomplete resolution and structural shift.}
Despite overall risk reduction, Table~\ref{tab2} shows that complete clearance remains limited: even the best-performing model (Opus 4.6) achieves only $R_c = 51.25\%$, meaning that nearly half of edited programs retain non-trivial vulnerabilities. This indicates that LLM-based edits primarily achieve partial mitigation rather than full resolution under weak-security constraints.

\begin{table}[t]
\centering
\caption{Risk reduction rates ($R_\infty$: worst-case, $R_0$: average-sensitive) and complete clearance rate ($R_c$) in \% for each model across four editing tasks. Best results per task are \textbf{bolded}.}
\label{tab2}
\footnotesize
\resizebox{\textwidth}{!}{%
\begin{tabular}{l*{5}{ccc}}
\toprule
\rowcolor{tabheader}
\textbf{Model} &
\multicolumn{3}{c}{\textbf{ADD}} & \multicolumn{3}{c}{\textbf{REMOVE}} & \multicolumn{3}{c}{\textbf{FIX}} & \multicolumn{3}{c}{\textbf{REFACTOR}} & \multicolumn{3}{c}{\textbf{TOTAL}} \\
\rowcolor{tabheader}
& $R_\infty$ & $R_0$ & $R_c$ & $R_\infty$ & $R_0$ & $R_c$ & $R_\infty$ & $R_0$ & $R_c$ & $R_\infty$ & $R_0$ & $R_c$ & $R_\infty$ & $R_0$ & $R_c$ \\
\midrule
\rowcolor{tabalt}
Opus 4.6               & \textbf{71} & \textbf{71} & \textbf{34} & \textbf{70} & \textbf{70} & 29 & \textbf{90} & \textbf{91} & 62 & \textbf{98} & \textbf{98} & 80 & \textbf{82.25} & \textbf{82.50} & \textbf{51.25} \\
GPT-5.4                & 66 & 66 & 29 & 67 & 67 & 32 & 89 & 89 & 59 & \textbf{98} & 97 & 82 & 80.00 & 79.75 & 50.50 \\
\rowcolor{tabalt}
Sonnet 4               & 67 & 67 & 31 & 68 & 68 & \textbf{37} & 87 & 87 & 55 & 95 & 95 & 79 & 79.25 & 79.25 & 50.50 \\
Deepseek-v4-flash      & 65 & 65 & 24 & 59 & 59 & 22 & 83 & 83 & \textbf{66} & 91 & 91 & \textbf{85} & 74.50 & 74.50 & 49.25 \\
\rowcolor{tabalt}
Kimi k2.5              & 61 & 61 & 21 & 64 & 64 & 22 & 72 & 72 & 43 & 90 & 90 & 73 & 71.75 & 71.75 & 39.75 \\
Haiku 4.5              & 54 & 54 & 8  & 50 & 47 & 15 & 76 & 76 & 61 & 84 & 84 & 70 & 66.00 & 65.25 & 38.50 \\
\rowcolor{tabalt}
Doubao-1.5-pro 32k     & 55 & 53 & 19 & 51 & 51 & 20 & 68 & 68 & 43 & 82 & 82 & 64 & 64.00 & 63.50 & 36.50 \\
GLM-4-32B              & 46 & 46 & 11 & 44 & 44 & 12 & 63 & 63 & 47 & 82 & 81 & 56 & 58.75 & 58.50 & 31.50 \\
\bottomrule
\end{tabular}}
\end{table}

Complementarily, Table~\ref{tab1} shows that stronger models induce higher $D_{\mathrm{TV}}$: Opus 4.6 achieves $\mu_{\mathrm{TV}} = 0.629$ compared to 0.455 for GLM-4-32B, indicating that more capable models restructure the vulnerability profile more substantially. However, since $D_{\mathrm{TV}}$ measures only the magnitude of distributional shift without distinguishing improvement from degradation, it should be interpreted alongside the directional metrics $R_0$ and $R_\infty$.

\begin{figure}[htbp]
\centering
\includegraphics[width=\textwidth]{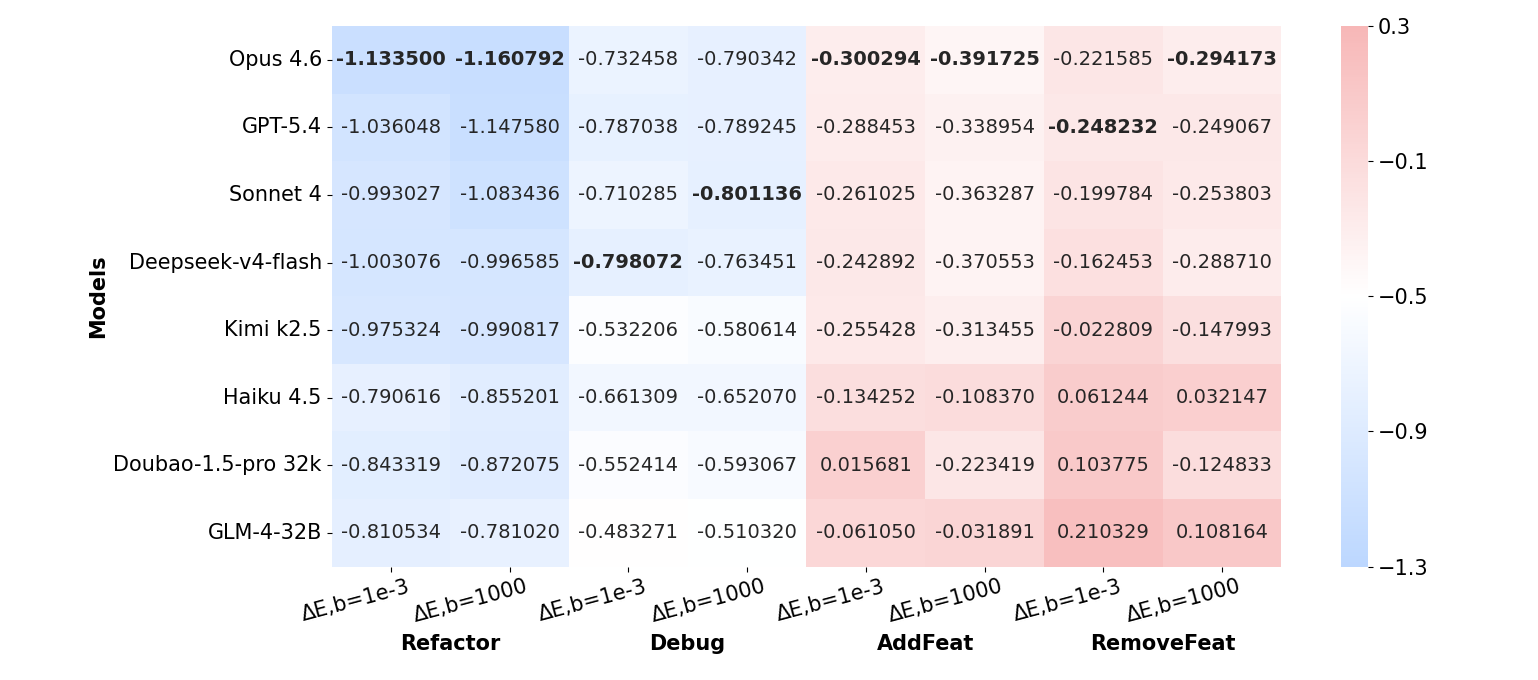}
\caption{Heatmap of per-task risk drift $\Delta E$ for eight models across four editing tasks under average-sensitive ($b{=}10^{-3}$) and worst-case-sensitive ($b{=}10^{3}$) regimes. Blue cells indicate risk reduction (negative drift); red cells indicate risk amplification (positive drift). REFACTOR and FIX consistently exhibit deeper blue across all models, while ADD occasionally produces near-zero or positive drift for weaker models.}
\label{fig_task}
\end{figure}
\vspace{-2mm}
\subsection{Task-Level Analysis}

We analyze how the four editing tasks affect security drift, focusing on reduction magnitude, worst-case sensitivity, and structural redistribution.

\paragraph{Task-level stratification.}
As shown in Table~\ref{tab2}, tasks form a clear two-group pattern. REFACTOR and FIX consistently outperform ADD and REMOVE across all metrics. Taking Opus 4.6 as an example, REFACTOR achieves $R_\infty{=}98\%$ and $R_c{=}80\%$, compared to only $R_\infty{=}71\%$ and $R_c{=}34\%$ for ADD---a gap of 27 and 46 percentage points, respectively. This pattern holds across models: even for the weakest model (GLM-4-32B), REFACTOR ($R_\infty{=}82\%$) substantially outperforms ADD ($R_\infty{=}46\%$).

We attribute this to the difference in transformation depth. REFACTOR and FIX involve structural reorganization or bug-level corrections that modify control flow more broadly, increasing the likelihood of incidentally altering vulnerability-related code paths. In contrast, ADD and REMOVE introduce or remove localized functionality, leaving existing vulnerability patterns largely intact.

\paragraph{Security drift across tasks.}
Fig.~\ref{fig_task} confirms this pattern from a drift perspective: FIX and REFACTOR induce more pronounced negative $\Delta E_0$ and $\Delta E_\infty$ across all models. Notably, for weaker models (e.g., GLM-4-32B, Doubao-1.5-pro), ADD occasionally produces near-zero or slightly positive $\Delta E_0$, suggesting that localized additions may introduce new insecure patterns or fail to remove existing ones.

\paragraph{Worst-case and structural effects.}
REFACTOR exhibits the strongest worst-case mitigation, with the largest reductions in $\Delta E_\infty$. This effect is amplified in stronger models, while weaker models show reduced separation between $\Delta E_0$ and $\Delta E_\infty$, suggesting limited ability to selectively mitigate high-severity vulnerabilities.

For distributional shifts, REFACTOR achieves the highest $D_{\mathrm{TV}}$ across models (e.g., $\mu_{\mathrm{TV}}{=}0.866$ for Opus 4.6), followed by FIX, while ADD and REMOVE yield consistently lower values ($\mu_{\mathrm{TV}} \approx 0.4$--$0.5$). This confirms that broader code transformations produce larger structural changes in vulnerability distributions.

\subsection{Case Study}

We illustrate a representative REFACTOR example where structural changes fail to eliminate the underlying vulnerability. The original code performs access control via \texttt{if not eval(client["scope\_check"], \{...\}, \{...\}):}, which directly evaluates externally influenced expressions---a classic code injection risk (CWE-94).

After refactoring by Kimi k2.5, the logic is reorganized into a class-based design with a dedicated \texttt{ScopePolicy.evaluate()} method, but the core \texttt{eval}-based execution of untrusted input remains unchanged.

This case produces near-zero $D_{\mathrm{TV}}$ and $\Delta E \approx 0$, illustrating that structural modernization does not necessarily translate into security improvement without explicit security instructions.

\subsection{Static and Dynamic Decomposition}
\label{sec:s_and_d}

\begin{table}[htbp]
\centering
\caption{Decomposition of average risk reduction $\overline{\Delta E_0}$ into static ($\overline{\Delta E_0^{(s)}}$) and dynamic ($\overline{\Delta E_0^{(d)}}$) components for each model. Both components contribute to overall risk reduction, with static reductions consistently larger.}
\label{tab:static_dynamic_split}
\small
\begin{tabular}{lccc}
\toprule
\rowcolor{tabheader}
\textbf{Model} & $\boldsymbol{\overline{\Delta E_0}}$ \textbf{(Total)} & $\boldsymbol{\overline{\Delta E_0^{(s)}}}$ \textbf{(Static)} & $\boldsymbol{\overline{\Delta E_0^{(d)}}}$ \textbf{(Dynamic)} \\
\midrule
\rowcolor{tabalt}
Opus 4.6           & -0.5969 & -0.4246 & -0.1723 \\
GPT-5.4            & -0.5899 & -0.4108 & -0.1792 \\
\rowcolor{tabalt}
Sonnet 4           & -0.5410 & -0.3902 & -0.1509 \\
Deepseek-v4-flash  & -0.5516 & -0.4121 & -0.1396 \\
\rowcolor{tabalt}
Kimi k2.5          & -0.4464 & -0.2611 & -0.1854 \\
Haiku 4.5          & -0.3812 & -0.2499 & -0.1313 \\
\rowcolor{tabalt}
Doubao-1.5-pro     & -0.3191 & -0.2003 & -0.1188 \\
GLM-4-32B          & -0.2861 & -0.1893 & -0.0968 \\
\bottomrule
\end{tabular}
\end{table}

Table~\ref{tab:static_dynamic_split} decomposes total risk reduction into static and dynamic components. Both consistently decrease, but static reductions are generally larger, suggesting that LLMs more effectively mitigate structural vulnerability patterns than execution-dependent issues.

\subsection{Fairness Study: Generation and Evaluation Independence}

To examine potential bias, we conduct a $2{\times}2$ controlled study on 40 seeds, crossing two generation pipelines (Claude Code vs.\ Cursor with DeepSeek-V4) with two evaluation models (Claude Opus vs.\ DeepSeek-V4). As shown in Table~\ref{tab:cross_fairness}, the variation across combinations is small, confirming that observed security drift is robust to both generation and evaluation choices.

\begin{table}[htbp]
\centering
\caption{Fairness analysis: average risk drift $\overline{\Delta E_0}$ under different generation--evaluation combinations.}
\label{tab:cross_fairness}
\small
\begin{tabular}{lcc}
\toprule
\rowcolor{tabheader}
& \textbf{Claude Gen} & \textbf{DeepSeek Gen} \\
\midrule
\rowcolor{tabalt}
Claude Eval     & -0.4410 & -0.4013 \\
DeepSeek Eval   & -0.3933 & -0.4127 \\
\bottomrule
\end{tabular}
\end{table}

\section{Limitations and Future Work}

\Name has several limitations. First, it focuses on programs of approximately 200 LOC, enabling controlled execution and systematic tracking of security signals under code transformations. However, this scale is much smaller than real-world software systems, which may limit generalization to large-scale and highly coupled codebases.

Second, we use Atheris for dynamic analysis, which captures input-driven crashes and certain runtime vulnerabilities, but may miss higher-level issues such as logic flaws and environment-dependent exploits. Thus, the dynamic component $E_b^{(d)}$ mainly reflects execution-level failure modes.

Third, under the weak-security setting, models are not explicitly instructed to consider security constraints and may rely on general security-related knowledge learned during pretraining when performing edits. This reflects realistic usage scenarios, but also makes it difficult to separate task-driven improvements from prior knowledge effects. Future work may explore stronger adversarial prompting and finer-grained control of security signals.

Finally, future work may extend \Name to larger and multi-module codebases and incorporate more comprehensive analysis tools.

\section{Conclusion}

We introduce \Name, a benchmark for studying security drift under LLM-driven code transformations in weak-security settings. By representing program security as a continuous attribute and defining drift measures over static and dynamic vulnerability signals, \Name enables fine-grained tracking of how risk evolves during functional code editing. Experiments on eight LLMs reveal consistent model stratification in security improvement capability, a bias toward worst-case mitigation, and task-dependent drift patterns linked to transformation depth---while also showing that no current model achieves full vulnerability resolution. \Name provides a systematic framework for transformation-aware security evaluation and supports future extensions to larger codebases and richer analysis tools.

%
%
%
\bibliographystyle{splncs04}
\bibliography{refer}

\end{document}